\documentclass[useAMS,usenatbib,letterpaper]{mn2e}
\usepackage{graphicx}
\usepackage{subfigure}

\def\kms      {\ifmmode{\rm km\,s}^{-1} \else km\,s$^{-1}$\fi}
\def\mujybm{\ifmmode{\rm \mu Jy}\,{\rm beam}^{-1}\else${\rm \mu}$Jy\,beam$^{-1}$\fi}
\def\ltsim{\ifmmode\stackrel{<}{_{\sim}}\else$\stackrel{<}{_{\sim}}$\fi}
\def\gtsim{\ifmmode\stackrel{>}{_{\sim}}\else$\stackrel{>}{_{\sim}}$\fi}
\def\farcs{\hbox{$.\!\!^{\prime\prime}$}}

\def\fsec{\hbox{$.\!\!^{\rm s}$}}
\def\hour{\hbox{$^{\rm h}$}}
\def\min{\hbox{$^{\rm m}$}}

\def\q24{q$_{\rm 24}$}

\title[Discovery of a new and unusual radio source in M82]{Discovery
  of an unusual new radio
  source in the star-forming galaxy M82: Faint supernova,
  supermassive blackhole, or an extra-galactic microquasar?}

\author[T. W. B. Muxlow et al.]{T. W. B. Muxlow$^{1}$, R. J. Beswick$^{1}$, S. T. Garrington$^{1}$,
  A. Pedlar$^{1}$, D. M. Fenech$^{2}$,\newauthor M. K. Argo$^{3}$, J. van Eymeren$^{1}$,
  M. Ward$^{4}$, A. Zezas$^{5,6}$, A. Brunthaler$^7$ \\
$^{1}$Jodrell Bank Centre for Astrophysics, School of Physics and
  Astronomy, The University of Manchester, Oxford Road, \\~~Manchester,
  M13 9PL\\
$^{2}$Department of Physics and Astronomy, University College London, Gower Street, London WC1E 6BT\\
$^{3}$ICRAR - Curtin Institute of Radio Astronomy, Curtin University of Technology, Bentley, Perth, WA 6845, Australia\\
$^{4}$Department of Physics, University of Durham, South Road, Durham, DH1 3LE\\
$^{5}$IESL, Foundation for Research and Technology, 71110 Heraklion, Crete, Greece\\
$^{6}$Harvard Centre for Astrophysics, 60 Garden Street, Cambridge,
  MA 02138, USA\\
$^7$Max-Planck-Institut f{\"u}r Radioastronomie, Auf dem H{\"u}gel 69, 53121 Bonn, Germany}

\begin{document}
\textheight 230mm
\topmargin -12mm
\date{Accepted 3rd March 2010; in original form 11th February 2010}

\pagerange{\pageref{firstpage}--\pageref{lastpage}} \pubyear{2010}

\maketitle

\label{firstpage}
\begin{abstract}
A faint new radio
source has been detected in the nuclear region of the starburst galaxy
M82 using MERLIN radio observations designed to monitor the flux
density evolution of the recent bright supernova SN\,2008iz. This new source was initially identified in observations made
between 1-5th May 2009 but had not been present in observations made
one week earlier, or in any previous observations of M82. In this
paper we report the discovery of this new source and monitoring of its
evolution over its first 9 months of existence. The true nature of
this new source remains unclear, and we discuss whether this
source may be an unusual and faint supernova, a supermassive blackhole
associated with the nucleus of M82, or intriguingly the first detection of radio
emission from an extragalactic microquasar.
\end{abstract}


\begin{center}
\begin{figure*}
\includegraphics[width=5.6cm,angle=270]{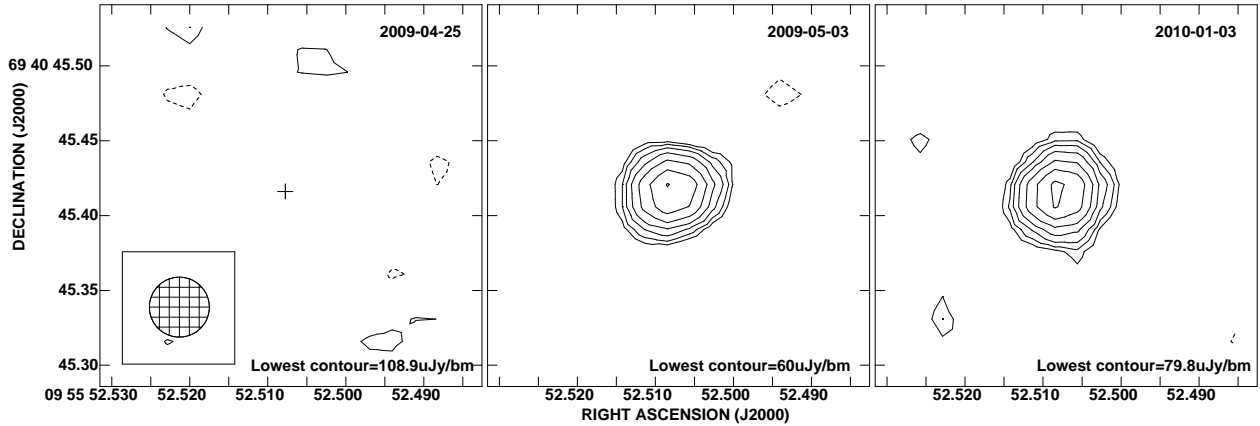}
\vskip -0.1cm
\caption{Images at the position of the new radio source in 3 epochs
  observed with MERLIN at 4994 (first two epochs) and 6668.4\,MHz (final
  epoch). All images have been convolved with a circular 40\,mas beam
  as shown in the bottom left of the first panel. The positional cross
 in first panel is the fitted position to Epoch 2
  (1-5th May 2009). Each image has been contoured in a similar manner
  with levels increasing by a factor of $\sqrt2$ times the lowest
  contour level, which is labeled in each panel.} 
\label{Maps1}
\end{figure*}
\end{center}
\vspace*{-1.6cm}
\section{Introduction}

The nearby \citep[d=3.6\,Mpc;][]{freedman94} star-forming galaxy M82 has been subject to
frequent radio monitoring at centimetric wavelengths with the VLA from
the early 1980s \citep{b1,b2}, and with MERLIN from the early 1990s
\citep{b3,b4}. Of order 60 compact radio sources have been
identified within the central kpc of M82, the majority of which are
thought to be recent supernova remnants which have exploded within the
last 2000 years. The origin of 46 of these objects has been determined
by the study of their radio spectral indices; 30 are considered to be
supernova remnants, and 16 are thought to be compact H{\sc ii} regions \citep{b5}. 

Radio monitoring at intervals of around a year has shown that there is
an additional population of radio transient sources whose origin is
unknown. To date two transient sources have been detected, and each
for only a single monitoring epoch implying that their lifetimes are
typically less than a year. \citet{b6} detected the compact radio
source 41.5+597 in M82 with the VLA in February 1981. At that epoch,
the object had a flux density of 7.1\,mJy and 2.6\,mJy at wavelengths
of 6 and 2\,cm respectively, implying that the source possessed a
steep radio spectral index of $\alpha$=$-$0.9 (S=$\nu$$^{\alpha}$). By
October 1983 the source had faded to below the detection threshold of
their VLA monitoring observations with an upper limit of 1.5\,mJy at
6\,cm. In a series of 6\,cm MERLIN monitoring observations starting in
the early 1990s, no emission was found at the position of 41.5+597 to
limits of $\sim$60$\mu$Jy, and in the deepest 6\,cm MERLIN
observations of M82 to date, made in 2002 \citep{b7} no emission was
found to a limit of 20\,$\mu$Jy.

In July 1992 \citet{b3} detected a second transient, 40.59+55.8 with
MERLIN at 6\,cm with a flux density of 1.2\,mJy. Subsequent MERLIN
monitoring at 21\,cm in April/May 1993 failed to detect emission at
the position of the transient to a limit of 300\,$\mu$Jy. Furthermore
it was not detected by deep MERLIN imaging at 6\,cm in February 1999
and April 2002 with limits of 35 and 21\,$\mu$Jy respectively
\citep{b5,b7}. Both 41.5+597 and 40.59+558 lie outside the dynamical
centre of M82 and since neither has given rise to a radio supernova
remnant, they may be examples of stellar binary microquasar
systems. If so, they would be the first to be discovered in the radio
outside the Milky Way.

Recently, \citet{brunthaler09a,b8} reported the detection of radio emission from a
new bright supernova in M82 (SN\,2008iz) which is thought to have flared
during the last week of March 2008 \citep{b10}. The appearance of
SN\,2008iz around 45 years after the previous supernova
\citep[43.31+592,][]{b11} is consistent with the radio supernova rate for M82 of a new
supernova approximately every 15 to 30 years \citep{b3,b7}. Subsequent
enhanced MERLIN monitoring of M82 has resulted in the detection of a
new radio source in the central region of the galaxy
\citep{b9}. This faint new radio source was discovered in observations
taken 1-5th May 2009 and was not present in images taken $\sim$1 week
earlier, on the 25th April 2009 (see Fig.\,\ref{Maps1}). Using closely spaced MERLIN (and VLBI) observations between
April 2009 and January 2010, it has been possible, for the first time, to
study the detailed evolution of one of the M82 transient source
population.

\vspace*{-0.5cm}
\section[]{Observations and Data Reduction}


The detection of both this new source and the campaign of continued
flux density monitoring of the evolving SN\,2008iz motivated a series of
radio monitoring observations of M82. MERLIN observations of M82 were
made between late April 2009 and January 2010  at 4994 and 6668.4\,MHz,
and 1658\,MHz.
 
 All observations were made in wide-field mode, with parallel hands of
 circular polarisation, measured over 16\,MHz of bandwidth correlated
 into 32 frequency channels. The primary flux density calibrator
 3C\,286 was used to set the flux density scale and the unresolved
 bright calibrator OQ208 was used to calibrate the amplitudes and
 bandpass responses. Throughout each epoch observations were
 interspersed with scans on the nearby phase reference source
 J095910+693217, with an assumed position of RA
 09$\hour\,59\min\,10\fsec6391$, Dec 69$\degr\,32\arcmin\,17\farcs723$
 (J2000).

Data from each epoch were independently reduced using standard methods
applying phase corrections determined form the phase reference
source, J095910+693217, and the data were weighted appropriately to account
for the relative sensitivities of the individual antennas. Following calibration, a large field encompassing
the entire radio extent of M82 at this resolution was imaged, using multiple
imaging facets and fully accounting for wide-field imaging effects. At each
different reference frequency all epochs were imaged in an identical
manner and the images were restored with a circular Gaussian beam
appropriate for the {\it uv} spacing of the baseline lengths and the
weighting applied to the gridded data during imaging.  

\begin{center}
\begin{figure*}
\includegraphics[width=15.5cm,angle=0]{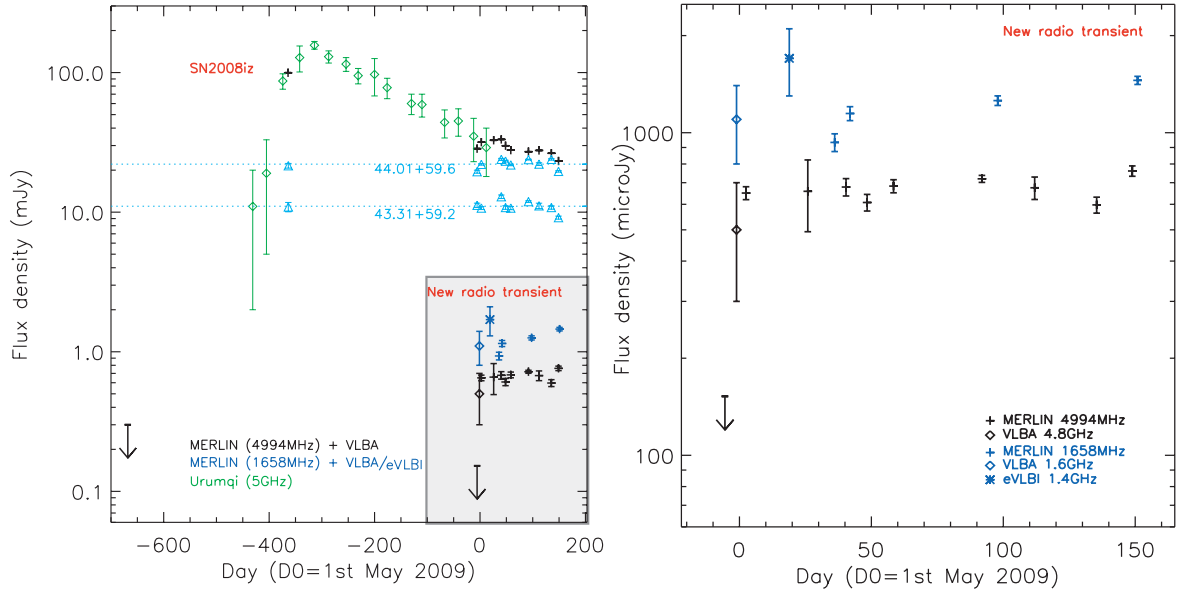}
\vskip -0.15cm
\caption{The {\it left-hand panel} shows the radio light curve of SN\,2008iz and the new transient source
  reported over the first 150 days of its existence. MERLIN 4994 and
  1658\,MHz data are plotted as black and blue crosses
  respectively. MERLIN observations at 6666.8\,MHz are not
  plotted. 1.4\,GHz eVLBI data are shown as blue stars, 5 and 1.6\,GHz
  VLBA are shown as black and blue open diamonds respectively
  \citep{brunthaler09d}. The 5\,GHz light curve for the SN\,2008iz
  \citep{b10} derived from single dish Urumqi observations is shown in
  green. The flux densities measured from these MERLIN data at 5\,GHz
  for two nearby compact remnants 44.01+59.6 and 43.31+59.2 which are
  known to have a constant flux density \citep{b6,ulvestad94} are also shown as
  pale blue triangles. An enlargement of the shaded region showing the
  light curve for the new radio transient is shown in the {\it
  right-hand panel}.}
\label{lightcurve1}
\end{figure*}
\end{center}
\vspace*{-1.0cm}
\section{Results}
\subsection{Radio lightcurve}

Radio lightcurves of the new MERLIN radio source are shown in
Fig.\,\ref{lightcurve1}, along with a composite MERLIN and Urumqi
5\,GHz light curve for SN\,2008iz \citep{b10,beswick09}. The Urumqi
5\,GHz observations do not resolve M82, but the observed variations in
the total flux density are dominated by SN2008iz. The new MERLIN radio
source reached a flux density of $\sim$600-700\,$\mu$Jy at 4994\,MHz
by early May 2009, showing greater than a factor of 5 increase in flux
density within 8 days. The flux density of the source has remained
approximately constant throughout all subsequent
observations. Simultaneous VLBA observations at 1.6 and 4.8\,GHz
observed on 30th April 2009, 3 days prior to the initial MERLIN 5\,GHz
detection, show this source to have an initial spectral index of
$-$0.7 \citep{brunthaler09d}. MERLIN monitoring observations at 1658,
4994 and 6668.4\,MHz show no significant variations in the spectral
index of this source throughout its first 150 days.

 The evolution of the radio light curve for SN\,2008iz, \citep[left-hand panel of
 Fig.\,\ref{lightcurve1} and discussed in detail by][]{b10}, is typical for a core-collapse
 supernova, showing a rapid rise in flux density followed by a
 power-law decline \citep[see for example][]{weiler02}. In comparison the new MERLIN
 radio source is $\sim$100 times fainter than SN\,2008iz, and shows
 significantly different flux density evolution with little or no
 detectable variation following a very rapid initial rise. 

\vspace*{-0.5cm}
\subsection{Position and size of this new radio source}
 
This new MERLIN source was detected on 3rd May 2009 at a position of
RA 09$\hour\,55\min\,52\fsec5083$, Dec
69$\degr\,40\arcmin\,45\farcs410$ (J2000) with an astrometric
error of 5\,mas in each coordinate. This position is within 3\,mas of
the VLBA detection of this source on the 30th April 2009 \citep{brunthaler09d}.

The position of the new MERLIN source has been measured in each epoch
relative to the position of the phase reference source and relative to
other bright, static radio sources in M82, such as SN\,2008iz,
41.95$+$57.5, 43.31$+$59.2 and 44.01$+$59.6. Over the first 50 days of
monitoring, including 6 MERLIN and 3 VLBI epochs, the fitted position
of the source shows evidence for east to west proper motion
of $\sim$10$\pm$5\,mas. This equates to an apparent proper motion of
$\sim$0.2\,mas\,day$^{-1}$, equivalent to an apparent superluminal
motion of $\sim$4.2c at the distance of M82. Subsequent data from 29th
June 2009 (58 day after 1st May 2009) onwards show the source position
to be consistent with its initial position measured on 3rd May
2009. Thus, considering that the positional shift is at the limit
achievable with these data the detection of any proper motion can only
be considered as tentative at this early epoch.

The highest resolution image with MERLIN was observed on 3rd January
2010, 247\,d after 1st May  2009, at a
frequency of 6.7\,GHz. From these data the source is partially
resolved with a deconvolved Gaussian fitted size of $15^{+5}\!\!\!\!\!\!_{-6}$\,mas.

\section {Discussion: nature of this new radio source}

Historical transients 41.5+597 \citep{b6} and 40.59+558 \citep{b3}
were each detected in only one monitoring epoch implying lifetimes of
less than a few months to a year. The new transient is broadly
similar to the earlier detections in flux density (within a factor of
10) and spectral index, although its longevity may soon indicate that
it is a different type of object. To date, beyond the radio detection
reported here, no confirmed detections of this new source has been
made at any other waveband in either archival or contemporaneous
observations, including X-ray \citep{kong09} and at K-band using
Gemini \citep{fraser09} and the Nordic Optical Telescope (S. Mattila,
private comm.).

We must consider the possibility that the new object is a background radio
source that has brightened significantly. The area of sky that has
been subjected to monitoring observations is the central nuclear
$\sim$1\,kpc of M82, an area of approximately 45$\times$15\,arcsec in
extent. The probability of finding a background AGN
system of $\sim$1\,mJy at 5\,GHz within this area is $\sim$1 in 550
\citep{prandoni06}. However, this object is extremely unusual in that it has
brightened by at least a factor of 5 (detection flux density/3$\sigma$
non-detection flux density) on a timescale of $\sim$1 week. Since
$<$1$\%$ of the faint background source population exhibit such violent intrinsic
variability, this reduces the probability by at least two orders
of magnitude \citep[e.g.][]{carilli03}. 

The position of this new source lies at high Galactic latitude
(+40$\degr$). Considering this and the fact that no detection of this new
source has been made in any other waveband this source is consistent with being either extremely
optically faint and/or highly obscured by material within M82. Thus on the balance of probability, we conclude that the new
transient is neither a foreground or background source and that it must
lie within M82. 

\begin{center}
\begin{figure}
\includegraphics[width=8.0cm]{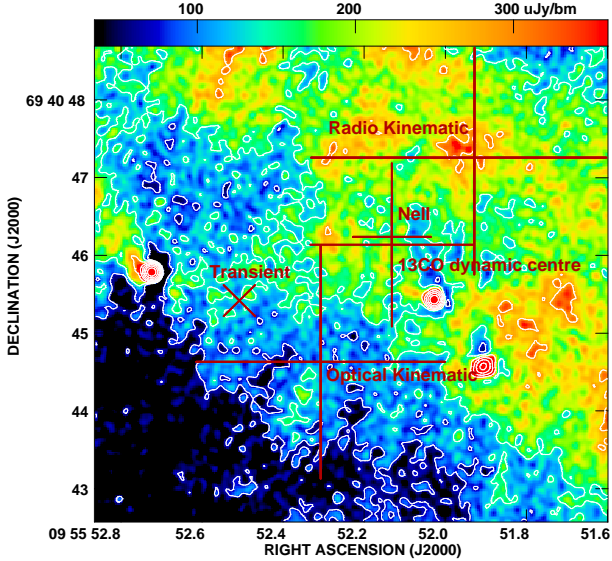}
\vskip -0.15cm
\caption{Combined MERLIN and VLA 5\,GHz false colour image of the region surrounding
  the new radio source location, taken prior to its discovery. The position of the new source is
  marked by an X (size not equal to the astrometric error). The position with associated errors of the dynamical centre of M82 derived
  via several multiwavelength methods (Weliachew et al.
  1984) are also shown as plus signs ($+$). The two brightest compact radio sources in this image are 44.01+59.6 (left) and 43.31+59.2 (right).}
\label{dyncentre}
 \end{figure}
\end{center}

\vspace*{-1.2cm}
\subsection{A faint and unusual radio supernova?}

Typically radio supernovae emit high brightness temperature
synchrotron emission which initially is absorbed at longer wavelengths. As the supernova shell
expands this absorption decreases resulting in a rapid turn-on of
emission at shorter wavelengths followed by a later turn-on at longer
wavelengths, with an associated evolution in radio
spectral index. After reaching its peak luminosity a supernova then normally follows
a power-law decline \citep[e.g. SN\,2008iz. Fig\,\ref{lightcurve1} and ][]{weiler02}. 

The peak luminosity of this new source at 5\,GHz is
$\sim$1$\times10^{18}$\,W\,Hz$^{-1}$. This is 3 orders of magnitude
less than the peak observed from Type-Ib/c supernovae and comparable
to the limits on the radio emission from Type-Ia
supernovae, which so far have not been detected \citep{panagia06}. Whilst this source is two orders of magnitude fainter
than  SN\,2008iz (see Fig.\,\ref{lightcurve1}), its luminosity is
comparable to some faint nearby Type-II radio
supernovae \citep[e.g. SN\,2004dj, SN\,1987A;][]{beswick05,turtle87}. The peak luminosity and the
rise time of Type-II supernovae can be empirically related \citep[see Eq. 20
  of][]{weiler02}. Following this relationship a
source of this luminosity should reach its peak flux density at 5\,GHz
 between 3 and 11\,days after the supernova detonation. This timescale is consistent with the rise time
observed (Fig.\,\ref{lightcurve1}) and  supports the scenario that
this source is a Type-II supernova.

However, there are several observational discrepancies with the
hypothesis that this is a faint supernova. Firstly, following the very
rapid initial rise in flux density (Figs.\,\ref{Maps1},
\ref{lightcurve1}) the light curve for this object shows no
significant evolution and in particular no power-law decline. Whilst
this is not common for supernovae \citep{weiler02}, a plateau in the
radio light curve as observed here could result from the expanding
supernova shell interacting with a denser interstellar medium at later
times. Thus, whilst atypical, this lack of power-law flux density
decay cannot rule out the supernova hypothesis. Secondly the spectral
index as measured via simultaneous multi-frequency observations prior
to the source reaching its peak flux density was $-$0.7 and has shown
no apparent evolution in the subsequent 150 days.  This characteristic
is in contrast to spectral evolution expected for young radio
supernovae, and can only be accounted for if the source had already
evolved past its peak at centimetre radio frequencies before 30th
April 2009.

The initial expansion velocities of typical radio supernovae have been
measured to be $\sim$23,000\,km\,s$^{-1}$ 
\citep[e.g. SN\,2008iz][and references therein]{brunthaler09c,brunthaler10,weiler02}. Thus at an age of
$\sim$250 days a radio supernova would typically have an angular size
of $\ltsim$2\,mas at the distance of M82. Our most recent and highest
resolution observations were taken on 3rd January 2010, when the
source was at least 250 days old. At this epoch the source is
tentatively resolved with a Gaussian fitted size of
$15^{+5}\!\!\!\!\!\!_{-6}$\,mas (see Section 3.2). If this source is a
supernova this size would require either a mildly relativistic
expansion velocity \citep[similar to that recently reported for
SN2007gr and SN2009bb,][]{paragi10,soderberg10} or that its age is
significantly underestimated.

\vspace*{-0.2cm}
\subsection{Accretion around a massive collapsed object?}

The steep radio spectral index from birth (and the possible detection
of apparent superluminal motion) supports the hypothesis that the
transient may be associated with an accretion disc around a
massive collapsed object in the nuclear region of M82. We suggest two possible scenarios.

\vspace*{-0.4cm}
\subsubsection{An AGN in the nucleus of M82}

The transient lies close to, but a few arcsec to the West of the
dynamical centre of M82 as derived from radio, optical, NeII, and
$^{13}$CO kinematic studies \citep[][see Fig.~3]{weliachew84}. The position is also displaced from the ridge-line
of the extended radio emission which is thought to be associated with
the integrated emission from ejected plasma over the recent
star-formation history of M82. Unless the region of nuclear
star-formation and the dynamical centre are significantly displaced
from the centre of the gravitational potential, it would seem unlikely
that this object is associated with a central super-massive black hole
(SMBH) in M82. Emission from such an object has, to date, never been
detected. It is possible that this could be emission from a system
associated with a second SMBH absorbed from a dwarf galaxy merging
with M82, but there is no supporting evidence of such a merger from
observations; however M82 is very disturbed and is interacting with
other galaxies in the M81-M82 group (Yun, Ho \& Lo 1994).

\vspace*{-0.3cm}
\subsubsection{Radio emission from an extragalactic microquasar?}

Alternatively this source may be result of
some form of flaring microquasar event in M82. The
700\,$\mu$Jy flux density of this source in M82 is equivalent to a
$\sim$90\,Jy source at a distance of 10\,kpc. The brightest
microquasar flares that have been seen in the Galaxy at centimetric
wavelengths are from Cygnus X-3 which flares to several tens of Jy
\citep{g1}. However, whilst such flares are close to the required
luminosity seen in the transient, the light curves of known Galactic
microquasars differ significantly from that seen for this object with
Galactic flares peaking and then decaying away on a timescale of days
to weeks. The radio luminosity of the transient is comparable with the
ultraluminous X-ray source found in the nearby dwarf galaxy NGC 5408
\citep{Lang}. Both objects possess a steep radio spectral index, however
to date, no variable X-ray source has been detected at this position in M82. 

Some form of relativistic jet could account for the observed steep
spectral index since in Galactic microquasars any optically thick
state typically evolves to thin within hours of turn on
\citep{Mirabel99,Fender04}. This scenario is compatible with the possible
detection of superluminal proper motion and the elongation seen at
late times. The strong jet-disk coupling would thus imply the presence
of an accretion disk around a massive collapsed object, although the
nature of this collapsed object remains unclear. If the transient is
some form of microquasar, its luminosity suggests that it is likely to
be associated with a massive black hole system of some type. This
could range from an extreme form of X-ray binary to an
intermediate-mass black hole system. However, the very high luminosity
and temporal longevity of the transient imply that this type of
accretion object is unusual and has not yet been seen within our
Galaxy.

\vspace*{-0.5cm}
\section{Conclusions}

This new source could be any of the above possibilities although each
of the proposed scenarios has difficulty in explaining all of the
observed properties. At present this source
has been detected for $>$9 months and shows no immediate signs of
fading. Depending upon its longevity it may represent
another example of a relatively short-lived faint radio source
population in M82. If so it would be the third example seen
in $\sim$30\,years of observations; thus a lower limit on their occurrence
rate is $\sim$1 every 10\,years, depending on their
lifetimes. If this population is associated with faint supernovae it will have significant implications on the radio derived
supernova rate of M82.  Alternatively if this source is associated
with microquasar flare event and the occurrence of which is related to
the host galaxies star-formation rate, we would expect to see a
comparable event in our own Galaxy around 1 every 100\,years. Regular
monitoring observations with new sensitive, high resolution imaging
arrays, such as e-MERLIN and the EVLA, will be required to determine
the size and nature of any such a population, both M82 and other
star-forming galaxies.

Global VLBI observations at 1.6 and 5\,GHz with milliarcsecond
resolution were taken in late 2009 and are awaiting
correlation. Images of the transient from these new data will
constrain the nature of this exciting source.

\vspace*{-0.7cm}
\section*{Acknowledgments}
MERLIN is operated by The University of Manchester
on behalf of the Science and Technology Facilities Council. 
We thank R. Spencer for useful discussion.


\label{lastpage}


\begin{thebibliography}{99}

\bibitem[\protect\citeauthoryear{Beswick et al.}{2005}]{beswick05}
Beswick, R.J., Muxlow, T.W.B., Argo, M. K., Pedlar, A., Marcaide,
J. M. \& Wills, K. A. 2005, ApJ, 623, 21
\bibitem[\protect\citeauthoryear{Beswick et al.}{2006}]{b11}
Beswick, R.J., Riley, J. D., Marti-Vidal, I.  et al., 2006, MNRAS, 369, 1221
\bibitem[\protect\citeauthoryear{Beswick et al.}{2009}]{beswick09}
Beswick, R.J., Muxlow, T.W.B., Pedlar, A., Fenech, D.M., Fender, R. \&
Maccarone, T., 2009, ATel, 2060
\bibitem[\protect\citeauthoryear{Brunthaler et al.}{2009a}]{brunthaler09a}
Brunthaler, A., Menten, K.M., Reid, M.J., Henkel, C., Bower, G.C. \& Falcke, H., 2009a, ATel, 2020
\bibitem[\protect\citeauthoryear{Brunthaler et al.}{2009b}]{b8}
Brunthaler, A., Menten, K.M., Reid, M.J., Henkel, C., Bower, G.C. \& Falcke, H., 2009b, A\&A, 499, L17
\bibitem[\protect\citeauthoryear{Brunthaler et al.}{2009c}]{brunthaler09c}
Brunthaler, A., Menten, K.M., Reid, M.J., Henkel, C., Bower, G.C.,
Falcke, H. 2009c, CBET, 1803
\bibitem[\protect\citeauthoryear{Brunthaler et al.}{2009d}]{brunthaler09d}
Brunthaler, A., Menten, K.M., Reid, M.J., Henkel, C., Bower, G.C. \& Falcke, H., 2009d, ATel, 2078
\bibitem[\protect\citeauthoryear{Brunthaler et al.}{2010}]{brunthaler10}
Brunthaler, A. et al, 2010 submitted to A\&A  
\bibitem[\protect\citeauthoryear{Carilli, Ivison \& Frail}{2003}]{carilli03}
Carilli, C.L, Ivion, R. J. \& Frail, D. A., 2003, ApJ, 590 192
\bibitem[\protect\citeauthoryear{Fender \& Belloni}{2004}]{Fender04}
Fender, R., \& Belloni, T., 2004, ARA\&A, 42, 317
\bibitem[\protect\citeauthoryear{Fenech et al.}{2008}]{b7}
Fenech, D.M., Muxlow, T.W.B., Beswick, R.J., Pedlar, A. \& Argo, M.K.,
2008, MNRAS, 391, 1384
\bibitem[\protect\citeauthoryear{Fraser et al.}{2009}]{fraser09}
Fraser, M., Smartt, S. J., Crockett, M., Matilla, S., Gal-Yun, A.,
Stephens, A. \& Roth, K., 2009, ATel, 2131
\bibitem[\protect\citeauthoryear{Freedman et al.}{1994}]{freedman94}
Freedman, W. L., Hughes, S. M., Madore, B. F., et al., 1994, ApJ, 427, 628
\bibitem[\protect\citeauthoryear{Gregory \& Kronberg}{1972}]{g1}
Gregory, P.C. \& Kronberg, P.P., 1972, Nature, 239, 440
\bibitem[\protect\citeauthoryear{Kong \& Chiang}{2009}]{kong09}
Kong, A. K. H. \& Chiang, Y.-K, 2009, ATel, 2080
\bibitem[\protect\citeauthoryear{Kronberg, Biermann \& Schwab}{1981}]{b1}
Kronberg, P.P., Biermann, P. \& Schwab, F.R., 1981, ApJ, 246, 751
\bibitem[\protect\citeauthoryear{Kronberg, Biermann \& Schwab}{1985}]{b2}
Kronberg, P.P., Biermann, P. \& Schwab, F.R., 1985, ApJ, 291, 693
\bibitem[\protect\citeauthoryear{Kronberg \& Sramek}{1985}]{b6}
Kronberg, P.P. \& Sramek, R.A., 1985, Science, 227, 28
\bibitem[\protect\citeauthoryear{Lang et al.}{2007}]{Lang}
Lang, C.C., Kaaret, P., Corbel, S., \& Mercer, A., 2007, ApJ, 666, 79
\bibitem[\protect\citeauthoryear{Marchili et al.}{2010}]{b10}
Marchili, N.,Marti-Vidal, I., Brunthaler, A.,  et al. 2010, A\&A, 509, 47 
\bibitem[\protect\citeauthoryear{McDonald et al.}{2002}]{b5}
McDonald, A.R., Muxlow, T.W.B., Wills, K., Pedlar, A. \& Beswick, R.J.,
2002, MNRAS, 334, 912
\bibitem[\protect\citeauthoryear{Mirabel \& Rodr{\`i}guez}{1999}]{Mirabel99}
Mirabel, I.F., \& Rodr{\`i}guez, L.F., 1999, ARA\&A, 37, 409
\bibitem[\protect\citeauthoryear{Muxlow et al.}{1994}]{b3}
Muxlow, T.W.B., Pedlar, A., Wilkinson, P.N.W., Axon, D.J., Sanders, E.M.,
\& de Bruyn, A.G., 1994, MNRAS, 266, 455
\bibitem[\protect\citeauthoryear{Muxlow et al.}{2009}]{b9}
Muxlow, T.W.B., Beswick, R.J., Pedlar, A., Fenech, D.M., Argo, M.K., Ward, M.J., \& Zezas, A.
2009, ATel 2073
\bibitem[\protect\citeauthoryear{Panagia et al.}{2006}]{panagia06}
Panagia, N., van Dyk, S. D., Weiler, K. W., Sramek, R. A., Stockdale,
C. J., \& Murata, K. P., 2006, ApJ, 646, 369
\bibitem[\protect\citeauthoryear{Paragi et al.}{2010}]{paragi10}
Paragi, Z., Taylor, G. B., Kouveliotou, C., et al, 2010, Nature, 463, 516
\bibitem[\protect\citeauthoryear{Pedlar \& Muxlow}{1995}]{b4}
Pedlar, A., \& Muxlow, T.W.B., 1995, Ap\& SS, 233, 281
\bibitem[\protect\citeauthoryear{Prandoni et al}{2006}]{prandoni06}
Prandoni, I., Parma, P., Wieringa, M. H., et al.,  2006, A\&A, 457, 517
\bibitem[\protect\citeauthoryear{Soderberg et al}{2010}]{soderberg10}
Soderberg, A. M., Chakraborti, S., Pignata, G., et al., 2010, Nature, 463, 513
\bibitem[\protect\citeauthoryear{Turtle et al.}{1987}]{turtle87}
Turtle, A. J., Campbell-Wilson, D., Bunton, J. D., Jauncey, D. L. \&
Kesteven, M. J., 1987, Nature, 327, 38
\bibitem[\protect\citeauthoryear{Ulvestad \& Antonucci}{1994}]{ulvestad94}
Ulvestad, J.S., \& Antonucci, R.R.J., 1994, ApJ, 424, L29
\bibitem[\protect\citeauthoryear{Weliachew, Fomalont \& Greisen}{1984}]{weliachew84}
Weliachew, L., Fomalont, E. B. \& Greisen, E. W., 1984 A\&A, 137, 335
\bibitem[\protect\citeauthoryear{Weiler et al.}{2002}]{weiler02}
Weiler, K. W., Panagia, N., Montes, M. J. \& Sramek, R. A.,
2002, ARA\&A, 40, 387
\bibitem[\protect\citeauthoryear{Yun, Ho \& Lo}{1994}]{yun94}
Yun, M. S., Ho, P. T. P. \& Lo, K. Y, 1994, Nature, 372, 530




\end{thebibliography}
\end{document}